# Optomechanical light storage in a silica microresonator

Victor Fiore, Chunhua Dong, Mark C. Kuzyk, and Hailin Wang

Department of Physics and Oregon Center for Optics
University of Oregon, Eugene, Oregon 97403, USA

Abstract

Coherent inter-conversion between an optical and a mechanical excitation in an optomechanical resonator can be used for the storage of an optical pulse as an excitation in a mechanical oscillator. This optomechanical light storage is enabled by external writing and readout pulses at one mechanical frequency below the optical resonance. In this paper, we expand an earlier experimental study [Phys. Rev. Lett. **107**, 133601 (2011)] on storing an optical pulse as a radial breathing mode in a silica microsphere. We show that the heterodyne beating between a readout pulse and the corresponding retrieved pulse features a periodic oscillation with a well-defined phase and with the beating period given by the mechanical frequency, demonstrating directly the coherent nature of the light storage process. The coherent inter-conversion accelerates with increasing optomechanical coupling rates, providing an effective mechanism for tailoring the temporal profile of the retrieved pulse. Experimental studies on both light storage and optomechanically induced transparency under nearly the same conditions also illustrate the connections between these two closely-related processes.

1) Introduction

Optical fields circulating in an optomechanical resonator can couple to the mechanical motion of the resonator via radiation pressure, leading to remarkable nonlinear optical processes[1,2]. Radiation-pressure-induced optical bistability has been observed in an early experimental study on cavity optomechanics[3]. Whereas optomechanical parametric instability was initially considered in the context of gravitational wave detectors[4-6], the experimental observation of radiation-pressure-induced coherent mechanical oscillations in silica microspheres and toroids ushered in the recent intense research efforts on micro- and nano-optomechanical resonators[1,2,7-9]. These studies have led to the experimental realization of a number of important and closely-related nonlinear optical phenomena, including strong coupling between an optical and a mechanical mode[10,11], coherent inter-conversion between optical and mechanical excitations[11,12], and optomechanically-induced transparency (OMIT)[13-15], which is the optomechanical analogue of the well-known phenomenon of electromagnetically induced transparency (EIT)[16].

The coherent inter-conversion between optical and mechanical excitations, which was proposed and analyzed in a number of earlier theoretical studies[17-21], has been used for the experimental realization of optomechanical light storage, storing an optical pulse as a mechanical excitation and retrieving the optical pulse from the mechanical oscillator at a later time[12]. Light stored as a mechanical excitation was first realized in an optical fiber via stimulated Brillouin scattering[22], though the mechanical lifetime in an optical fiber is limited to only a few nanoseconds. The use of mechanical oscillators with ultrahigh quality factors can enable a storage lifetime exceeding a millisecond. In comparison with atomic or spin systems that have been widely used for light storage[23], optomechanical light storage processes can not only store light at a given wavelength, but also map the stored mechanical excitation back to light at other desired wavelengths[20,21,24]. This capability of wavelength conversion can play a special role in a hybrid quantum network, by mapping photons emitted from one type of quantum system to photons that can couple to another type of quantum system. Although a single optomechanical resonator can be used for the storage of only a single optical mode, the spatial-temporal profile of an optical pulse can in principle be stored in an array of optomechanical resonators via processes analogous to dark-state polaritons in atomic media[25].

In this paper, we report experimental studies of optomechanical light storage, expanding our earlier study on storing an optical pulse as a radial breathing mode in a silica microsphere[12].

Studies of the heterodyne beating between a retrieved pulse and the corresponding readout pulse demonstrate directly the coherent nature of the light storage process. These studies also show that varying the inter-conversion rate between the optical and the mechanical excitations effectively tailors the temporal profile of the retrieved pulse. In atomic systems, the light storage process is closely related to EIT. Comparisons between light storage and OMIT under nearly the same experimental conditions illustrate similar connections between these two processes in an optomechanical system.

2) Theoretical model

For a Fabry-Perot resonator with one of the mirrors serving as the mechanical oscillator (see Fig. 1a), the mechanical displacement of the mirror modulates the frequency of the cavity resonance, with $\omega_c(x) = \omega_0 + g_{om}x$, where $x$ is the mechanical displacement and $g_{om}$ is the optomechanical coupling coefficient. Here, we consider the optomechanical coupling between an optical cavity mode and a mechanical oscillator, with the coupling driven by a strong external laser field at the red sideband, i.e., one mechanical frequency, $\omega_m$, below the cavity resonance, $\omega_0$, as shown schematically in Fig. 1b. In the resolved-sideband limit, the linearized optomechanical Hamiltonian is given by

$$H = \hbar\omega_m \hat{b}^+\hat{b} - \hbar\Delta \hat{a}^+\hat{a} + \hbar G(\hat{a}^+\hat{b} + \hat{a}\hat{b}^+) \qquad (1)$$

where $\hat{b}$ is the annihilation operator for the mechanical mode, $\hat{a}$ is the annihilation operator for the intracavity signal field in the rotating frame of the external driving field, $\Delta$ is the detuning between the external driving field and the optical cavity resonance, and $G = g_{om}x_{zpf}\sqrt{n_c}$ is the effective optomechanical coupling rate with $x_{zpf}$ being the zero-point fluctuation for the mechanical oscillator and $n_c$ being the intracavity photon number of the driving field. As shown in Eq. 1, the external driving field couples the signal field to the mechanical displacement, which can induce a coherent inter-conversion between the optical and motional states.

For optomechanical light storage, external "writing" and "readout" laser pulses at the red sideband drive the inter-conversion between the optical signal field and the mechanical excitation. For the storage process, the writing pulse couples a signal pulse, which is resonant with the optical cavity resonance, to the mechanical oscillator, generating a coherent mechanical excitation. For the retrieval process, a readout pulse interacts with the stored mechanical excitation, converting the mechanical excitation back to an optical pulse at the cavity resonance.

Similarly for OMIT, a control field at the red sideband couples to a signal field at the cavity resonance, generating a coherent mechanical excitation through the inter-conversion process discussed above. Anti-Stokes scattering of the control field off this mechanical excitation then generates an intracavity optical field at the cavity resonance. Destructive interference between the intracavity field generated by the anti-Stokes scattering and that generated by the input signal prevents the excitation of the optical mode, leading to OMIT[26].

With optical and mechanical damping processes included, the dynamics of the optomechanical system excited by an input field nearly resonant with the optical cavity resonance can be described by the following equations of motion:

$$\dot{\alpha} = [i(\omega_{in} - \omega_0) - \frac{\kappa}{2}]\alpha - iG\beta + \sqrt{\kappa^{ext}} A_{in}$$
$$\dot{\beta} = [i(\omega_{in} - \omega_0 - \Delta - \omega_m) - \frac{\gamma_m}{2}]\beta - iG\alpha$$

(2)

where $\omega_{in}$ and $A_{in}$ are the frequency and amplitude of the input optical signal field, respectively ($A_{in}$ is normalized such that $I_{in} = |A_{in}|^2$ is the input photon flux), $\kappa$ and $\kappa^{ext}$ are the total cavity decay rate and the effective output coupling rate of the signal field, respectively, $\gamma_m$ is the mechanical damping rate, and we define $\alpha = <\hat{a}> \exp[i(\omega_{in} - \omega_0 - \Delta)t]$ and $\beta = <\hat{b}> \exp[i(\omega_{in} - \omega_0 - \Delta)t]$. As will be discussed in detail later, both the OMIT and the optomechanical light storage can be described by the coupled oscillator equations given in Eq. 2.

3) Samples and experimental setup

Silica microspheres with a diameter near 30 μm were fabricated from an optical fiber tip with a focused $CO_2$ laser beam[27]. The microsphere was attached to a fiber stem with a diameter near 5 μm. Whispering gallery modes (WGMs) in a microsphere were excited through the evanescent field of a tapered optical fiber. The microsphere was positioned relative to the tapered fiber with a piezo-driven nano-positioning system. The microsphere and the tapered fiber were kept under a helium atmosphere to prevent contamination from moisture or airborne dust particles. The (1, 2) mechanical breathing mode of the microsphere was used in the light storage experiment. The (1, 2) mode has a mechanical quality factor considerably greater than that of the pure radial breathing (1, 0) mode[28-30]. Two samples were used for the experimental results presented here. For sample A, which was used for Fig. 3, ($\omega_m, \gamma_m, \kappa$)/2π = (160, 0.013, 6) MHz. For sample B, which was used for Figs. 4 and 5, ($\omega_m, \gamma_m, \kappa$)/2π = (160.9, 0.096, 20) MHz.

Figure 2 shows a diagram of the experimental setup that enabled us to carry out studies of both light storage and OMIT under the same or similar experimental conditions. The writing, readout, and signal pulses with wavelengths near 800 nm were derived from a single-frequency Ti:Sapphire ring laser. A combination of acousto-optic modulators (AOMs) and electro-optic modulators (EOMs) were used to generate optical pulses with the desired duration, timing, and frequencies. Specifically, the writing and read-out pulses at the red sideband of the cavity resonance were derived from the first order diffraction from $AOM_1$. The signal pulse was derived from the blue sideband generated by passing the writing pulse through $EOM_1$. Unless otherwise specified, the writing and the signal have the same timing and duration. A Pound-Drever-Hall (PDH) frequency stabilization technique was used to lock the frequency of the ring laser relative to the given WGM of the microsphere. The locking pulse was generated by sending the zeroth order diffraction off $AOM_1$ through $AOM_{lock}$ (with a double pass configuration) and then through $EOM_{lock}$. The locking pulse arrived at the microsphere a few tens of μs before the signal pulse. Figure 1c illustrates the timing of the optical pulses.

Light emitted from the microsphere was measured with heterodyne detection. A spectrum analyzer in a time-gated detection mode was used for measuring the power of the heterodyne beats as a function of time, with the time resolution set by the resolution bandwidth (RBW) as well as the gate length (see Fig. 1c). Details of the gated-heterodyne-detection have been discussed in an earlier study[12]. For the light storage experiment, the writing and readout pulses served as the local oscillator for the heterodyne detection of the signal and retrieved pulses emitted from the microsphere, respectively. Similarly for the OMIT experiment, the control pulse at the red sideband served as the local oscillator for the signal pulse emitted from the microsphere. As discussed in the earlier study[12], the heterodyne detection is not sensitive to the part of the signal pulse that is not emitted from the optical resonator, since the input signal pulse is generated directly from the local oscillator with electro-optic phase modulation. In this regard, the emission power measured is directly proportional to the power of the intracavity optical field.

4) Experimental results

We first present experimental results on optomechanical light storage. Figure 3a shows the power of the signal and retrieved pulses emitted from the microsphere as a function of time, obtained with time-gated-heterodyne detection and with a fixed delay of 8 μs between the writing and readout. The duration of the writing and readout pulses is 1 μs. The incident peak power of the writing and readout pulses is $P_w = P_r = 6$ mW, corresponding to an estimated optomechanical coupling rate, $G/2\pi = 0.77$ MHz (as will be discussed in detail in Section 5, OMIT measurements were used to determine the effective optomechanical coupling rate). The input signal pulse is at the cavity resonance, with an incident peak power $P_{sig} = 0.1$ mW. A gate length of 100 ns was used, with the RBW set to 30 MHz. The result shown in Fig. 3a is a significant improvement in time resolution over our earlier work that used RBW= 1 MHz and a detection gate length of 3 μs. As investigated in detail in the earlier study, the energy of the retrieved pulse decreases with increasing separation between the writing and readout pulse (not shown), reflecting the decay of the coherent mechanical excitation induced by the signal and the writing pulse[12]. The storage lifetime is determined by the damping time of the mechanical oscillator.

We further characterized the retrieved pulse by measuring directly in the time domain the heterodyne beating between the retrieved and the readout pulse, more specifically by time-resolving the heterodyne beat signal with a digital oscilloscope. Figure 4 shows the heterodyne beats as a function of time, obtained at three different incident powers for the readout, with $P_r = 0.3$, 1.6, 2.9 mW, corresponding to $G/2\pi = 0.2$, 0.45, 0.6 MHz, respectively. The duration of the writing and readout pulses is respectively 1 μs and 6 μs. The relatively long readout pulse allowed us to probe how the temporal behavior of the retrieved pulse depends on the optomechanical coupling rate or the inter-conversion rate between the optical and the mechanical excitations. Other parameters used include $P_w = 1.6$ mW and $P_{sig} = 0.1$ mW. With increasing $P_r$, the mechanical excitation is converted to an optical field at a faster rate, leading to an effectively shorter duration for the retrieved pulse, as will be discussed in more detail in Section 5.

Figure 4b plots a portion of the heterodyne beats in Fig. 4a with an expanded timescale and also compares the experimental result with a periodic oscillation with a frequency given by the mechanical oscillator. Because of the limited data size, there are only three data points in a given oscillation period. Nevertheless, the excellent agreement between the experiment and the periodic oscillation shows that the heterodyne beats are characterized by a periodic oscillation with a well-defined phase and with the beat period given by the mechanical frequency, which confirms the coherent nature of the storage and retrieval processes.

We now present the experimental results on OMIT, obtained with the same setup as the light storage experiment. A relatively long writing pulse (8 μs in duration) served as the control for the OMIT experiment. The top trace in Fig. 5a shows the power of the signal field emitted from the microsphere as a function of the detuning between the signal field and the control field, with the control fixed at the red sideband. The emission spectrum was obtained with time-gated-heterodyne detection, with the detection gate set at 6.5 μs from the start of the writing pulse (see the inset of Fig. 5a) and with a gate length of 1 μs. The timing of the detection gate and the relatively long control pulse enabled us to probe the steady state OMIT process. Other parameters used include $P_{sig}$= 0.1 mW and $P_w$= 1.2 mW, corresponding to an estimated $G/2\pi$= 0.38 MHz. A pronounced sharp dip is observed in the emission spectrum when the detuning equals the frequency of the mechanical oscillator. Since the emission power measured is directly proportional to the power of the intra-cavity signal field, the dip at the anti-Stokes resonance corresponds to the suppression of the optical excitation induced by the OMIT.

For comparison, the bottom trace of Fig. 5a shows the power of the retrieved pulse emitted from the microsphere as a function of the detuning between the signal field and the writing field, obtained with otherwise the same condition as the top trace except for the application of a readout pulse with a pulse duration of 3 μs and with $P_r$= 1 mW. The detection gate with a length of 1 μs was set at the center of the readout pulse. The expanded plot of the emission spectra near the anti-Stokes resonance shown in Fig. 5b indicates that the OMIT dip (top trace) closely corresponds to the anti-Stokes resonance for the light storage process (bottom trace). Detailed theoretical analysis will be presented in Section 5.

5) Analysis and discussion

The experimental results presented in Section 4 are in good agreement with the theoretical calculations based on the use of the coupled oscillator equations shown in Eq. 2. Figure 3b plots the theoretically-calculated temporal profile of the signal and retrieved pulses emitted from the resonator, with all parameters determined from the experiments. The figure also plots the intensity of the coherent mechanical excitation induced by the storage process, illustrating the generation, decay, and then conversion of the mechanical excitation. A comparison between the experiment and theory shown in Figs. 3a and 3b indicates that even with the much improved time resolution, the time-gated detection still caused significant distortion in the temporal profile of the retrieved pulse.

The temporal profile of the retrieved pulse can be determined from a direct time-domain measurement of the heterodyne beating between the retrieved pulse and the corresponding readout pulse, as shown in Fig. 4. For a theoretical analysis, Fig. 4 also plots the calculated amplitude of the retrieved pulse as a function of time. Except for an overall scale factor for the amplitude, all parameters used were determined from the experiment, indicating good agreement between the experiment and theory. With increasing readout power, the increase in the corresponding optomechanical coupling rate leads to a faster conversion from the mechanical to the optical excitation and thus a shorter duration for the retrieved pulse. In this regard, the readout process can be used not only for the retrieval of the stored optical excitation, but also for pulse shaping, as suggested in earlier theoretical studies.

The OMIT spectrum shown in Fig. 5a is in good agreement with the theoretically calculated spectrum of the intra-cavity power as a function of the detuning between the signal and the control. For the theoretical calculation, the optomechanical cooperativity, $C = 4G^2/\gamma_m \kappa$, was used as a fitting parameter, with no other adjustable parameters. The optomechanical coupling rates, including those given in other figures, were then determined from the cooperativity. The theoretical calculation describes well both the width and the depth of the OMIT dip (see also Fig. 5b, which features an expanded frequency scale). The spectral response of the light storage process (i.e. the power of the retrieved pulse as a function of the detuning between the signal and the writing pulse) also shows good agreement between the experiment and the theoretical calculation using the same set of parameters as those for the OMIT experiment (see Fig. 5b).

Optomechanical light storage is closely related to OMIT. For both processes, the writing (or control) field at the red sideband interacts with the signal field at the cavity resonance via optomechanical coupling, generating a coherent mechanical excitation. For the OMIT process, the control field scatters off the mechanical excitation, leading to destructive interference between the signal field and the optical field generated from the anti-Stokes scattering process. For the transient light storage process, a readout pulse, arriving at a later time, scatters off the mechanical excitation, generating the retrieved pulse. In this case, there is no or little temporal overlap between the signal field and the optical field generated from the anti-Stokes scattering of the readout pulse. For comparison between light storage and OMIT, we have used in Fig. 5 a long writing pulse as the control for OMIT and have also used time-gated detection to probe the steady-state OMIT response. This leads to similar linewidth for the spectral responses of the OMIT and light storage processes. In general, for a steady-state OMIT process, the width of the OMIT dip is given by $(1+C)\gamma_m$, whereas for a light storage process, the width of the spectral response is determined by the duration of the writing pulse as well as $\gamma_m$ and $C$.

6) Conclusion

In summary, we have carried out detailed experimental studies on optomechanical light storage. These studies demonstrate the coherent nature of the light storage process and show the dependence of the temporal profile of the retrieved pulse on the optomechanical coupling rate. We have also compared the optomechanical light storage process to the closely related process of OMIT. A straightforward extension of the light storage process discussed here is to use a readout pulse that couples to a different optical mode. The resulting readout process enables optical wavelength or optical mode conversion[31, 32].

The inter-conversion between optical and mechanical excitations is limited by the thermal excitation of the mechanical oscillator. Effects of the thermal mechanical motion, which can be made negligible for classical applications, can prevent the application of the inter-conversion process in a quantum regime. The thermal mechanical motion can be eliminated by cooling the mechanical oscillator to its quantum ground state, as demonstrated in a number of recent experimental studies[11, 33-35]. Effects of thermal mechanical motion can also be circumvented with the use of an optomechanical dark mode[36-38], as recently realized by coupling two, instead of one, optical modes to a mechanical oscillator[39].

This work is supported by the DARPA-MTO ORCHID program through a grant from AFOSR and by NSF.

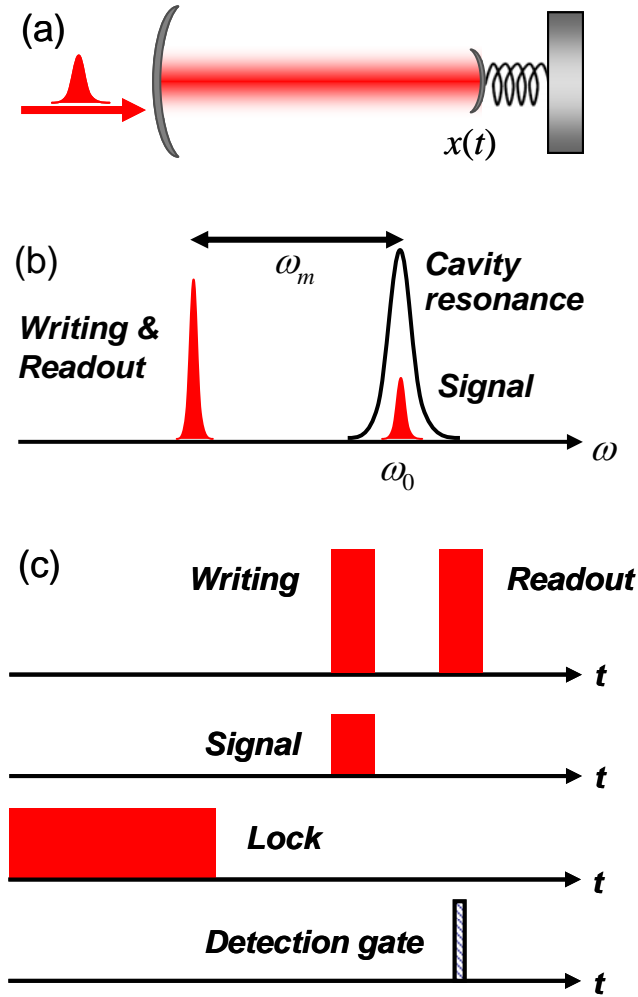

FIG 1. (Color online) (a) Schematic of an optomechanical resonator. (b) Spectral position for the writing, readout, and signal pulses used in optomechanical light storage. (c) Schematic of the pulse sequence and the detection gate used for the light storage experiments.

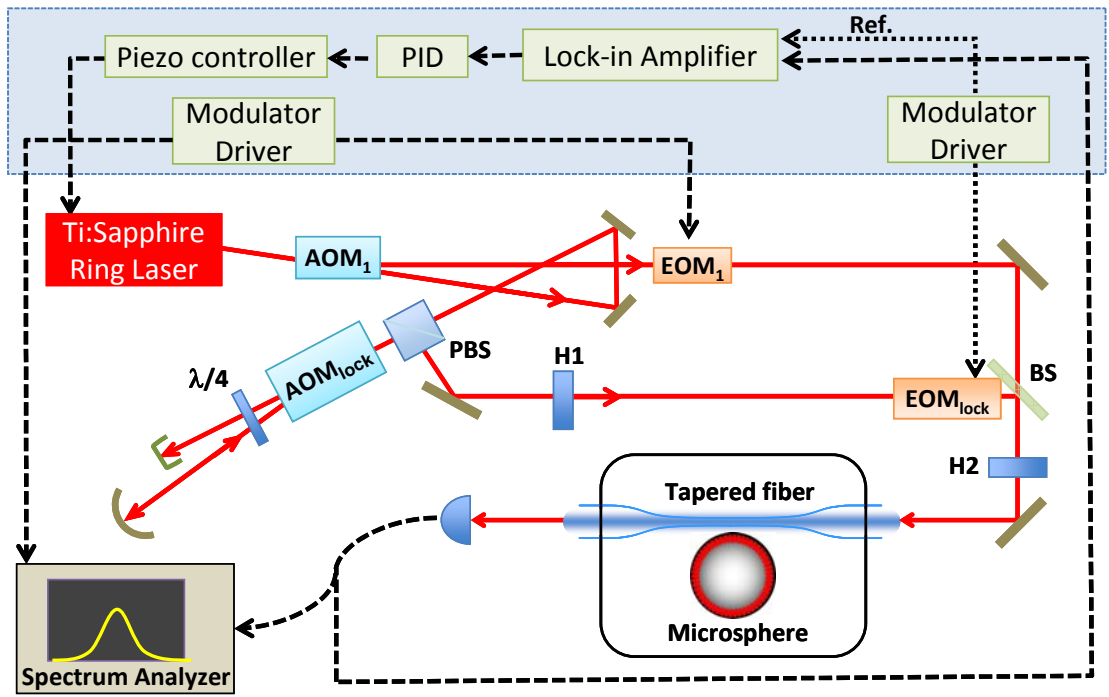

FIG 2. (Color online) Schematic of the experimental setup for both light storage and OMIT, with solid lines representing the optical paths and dashed lines representing the electrical connections. PBS, BS, λ/4, H1, and H2 are polarization beam splitter, beam splitter, quarter wave plate, half-wave plate 1, and half wave plate 2, respectively.

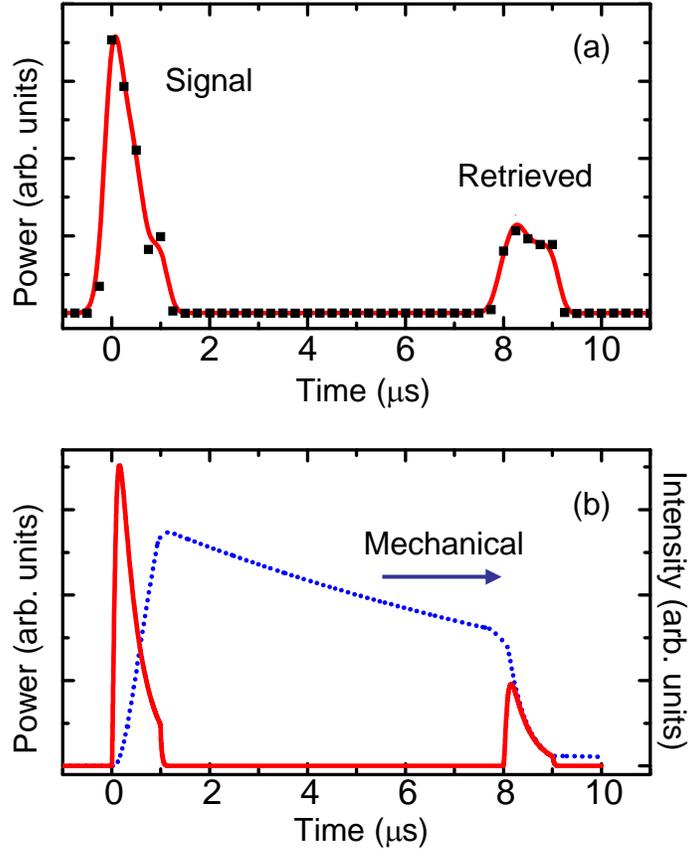

FIG. 3. (Color online) (a) The power of the signal and retrieved pulse emitted from the microsphere and measured with time-gated-heterodyne detection and with an 8 μs delay between the writing and readout. Other parameters are given in the text. The solid curve is a guide to the eye. (b) Theoretically calculated intra-cavity power of the signal and retrieved pulse (solid curve) and the intensity of the mechanical excitation involved in the light storage process (dashed curve), as discussed in the text.

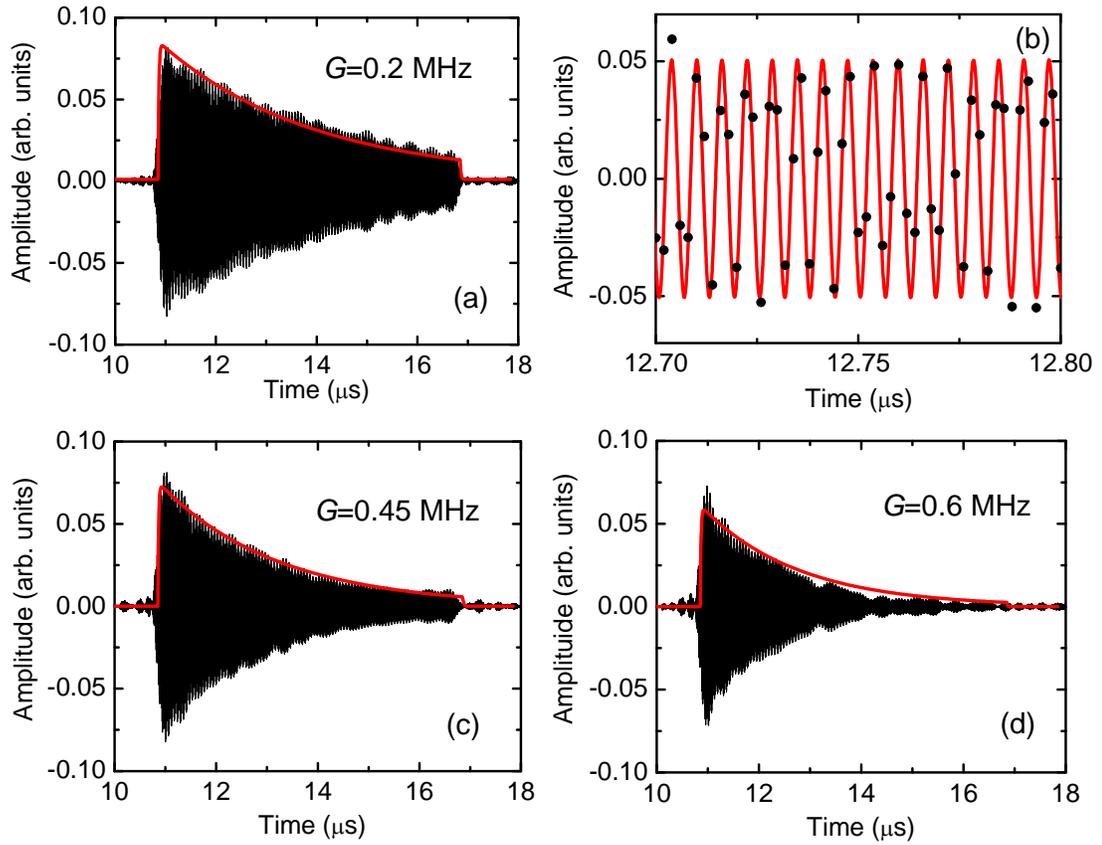

FIG. 4. (Color online) The heterodyne-beating between the retrieved pulse and the readout pulse. (b) shows a portion of the beats in (a), with an expanded timescale and with the solid curve being a periodic oscillation with a frequency given by the mechanical oscillator. The optomechanical coupling rate of the readout pulse is indicated in (a), (c), and (d). Other parameters are given in the text. The pulse envelope (red curve) shown in (a), (c), and (d) are the theoretically calculated amplitude for the retrieved pulse, as discussed in the text.

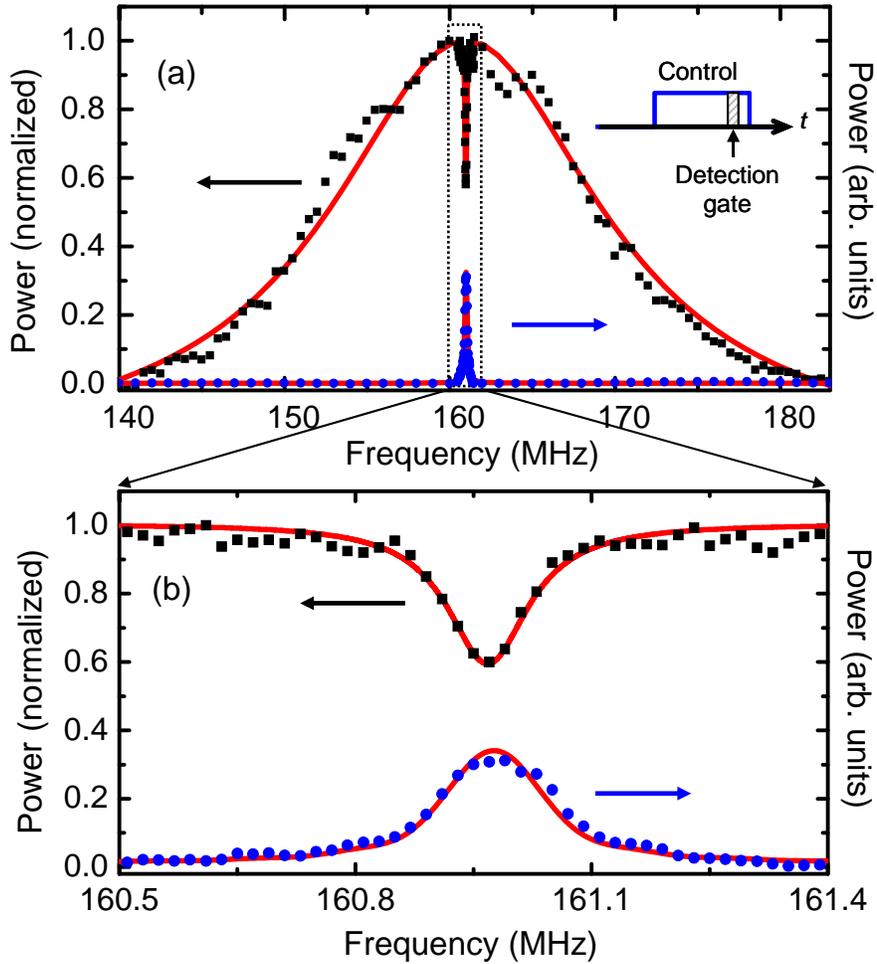

FIG. 5. (Color online) (a) Spectral response of OMIT (top trace) and light storage (bottom trace). (b) The spectral response in (a) near the anti-Stokes resonance, with an expanded frequency scale. Squares in the top trace are the signal power emitted from the microsphere. Circles in the bottom trace are the power of the retrieved pulse emitted from the microsphere. Both are obtained as a function of the detuning between the signal field and the control field, which is fixed at the red sideband. Solid curves are the theoretically calculated spectral response, as discussed in the text. The inset in (a) shows the timing of the detection gate used for the OMIT experiment.